\documentclass{article}
\usepackage{arxiv}

\usepackage[utf8]{inputenc} 
\usepackage[T1]{fontenc}    
\usepackage{hyperref}       
\usepackage{url}            
\usepackage{booktabs}       
\usepackage{amsfonts}       
\usepackage{nicefrac}       
\usepackage{microtype}      
\usepackage{lipsum}
\usepackage{fancyhdr}       
\usepackage{graphicx}       
\usepackage{multirow} 
\graphicspath{{figures/}}     
\usepackage{CJKutf8}
\usepackage{tikz}
\usepackage{tabularx}
\usepackage{array}
\usepackage{algorithm}
\usepackage{algpseudocode} 
\usepackage{amsmath}     
\usepackage[utf8]{inputenc}
\usepackage{booktabs} 
\usepackage{tabularx} 
\usepackage{enumitem} 
\usepackage{caption}  
\usepackage{float}
\usepackage[justification=centering]{caption} 
\usepackage{subcaption} 
\usepackage{pgfplots} 
\pgfplotsset{compat=1.18} 
\usepackage[section]{placeins} 

\PassOptionsToPackage{numbers, compress}{natbib}

\makeatletter

\title{ACE-Step 1.5: Pushing the Boundaries of Open-Source Music Generation}

\author{
Junmin Gong\(^{*}\) \\
\texttt{junmin@acestudio.ai} \\
ACE Studio \\
\and
Yulin Song\(\dagger\) \\
\texttt{keylxiao@acestudio.ai} \\
ACE Studio \\
\and
Wenxiao Zhao \(\dagger\) \\
\texttt{sean@acestudio.ai} \\
ACE Studio \\
\and
Sen Wang \(\dagger\) \\
\texttt{sayo@acestudio.ai} \\
ACE Studio \\
\and
Shengyuan Xu \(\dagger\) \\
\texttt{shengyuan@acestudio.ai} \\
ACE Studio \\
\and
Jing Guo\(\dagger\) \\
\texttt{joe@acestudio.ai} \\
ACE Studio \\
\and
Xuerui Yang \(\dagger\) \\
\texttt{yangxuerui@stepfun.com} \\
Step Fun \\
}

\begin{document}
\maketitle
\vspace{-10.5cm} 
\begin{center}
\includegraphics[width=0.65\linewidth]{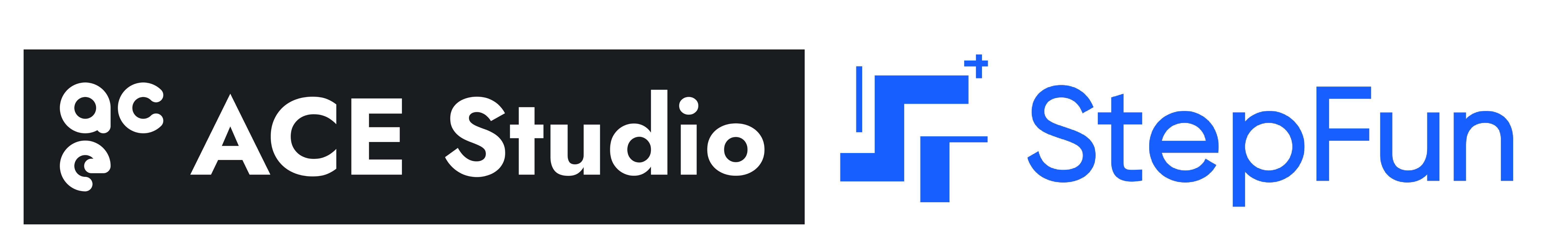}
\end{center}
\vspace{8cm} 
\begin{abstract}
We present ACE-Step v1.5, a highly efficient open-source music foundation model that brings commercial-grade generation to consumer hardware. On commonly used evaluation metrics, ACE-Step v1.5 achieves quality beyond most commercial music models while remaining extremely fast—under 2 seconds per full song on an A100 and under 10 seconds on an RTX 3090. The model runs locally with less than 4GB of VRAM, and supports lightweight personalization: users can train a LoRA from just a few songs to capture their own style. At its core lies a novel hybrid architecture where the Language Model (LM) functions as an omni-capable planner: it transforms simple user queries into comprehensive song blueprints—scaling from short loops to 10-minute compositions—while synthesizing metadata, lyrics, and captions via Chain-of-Thought to guide the Diffusion Transformer (DiT). Uniquely, this alignment is achieved through intrinsic reinforcement learning relying solely on the model's internal mechanisms, thereby eliminating the biases inherent in external reward models or human preferences. Beyond standard synthesis, ACE-Step v1.5 unifies precise stylistic control with versatile editing capabilities—such as cover generation, repainting, and vocal-to-BGM conversion—while maintaining strict adherence to prompts across 50+ languages. This paves the way for powerful tools that seamlessly integrate into the creative workflows of music artists, producers, and content creators. The code, the model weights and the demo are available at: \href{https://ace-step.github.io/ace-step-v1.5.github.io/}{https://ace-step.github.io/ace-step-v1.5.github.io/}.
\end{abstract}
\section{Introduction}
\label{sec:intro}

The domain of neural music generation has undergone a paradigm shift, with proprietary systems demonstrating remarkable fidelity and structural coherence. However, a significant disparity remains between closed-source capabilities and open-source alternatives. While our preceding iteration, ACE-Step v1.0~\cite{gong2025acestepstepmusicgeneration}, validated the feasibility of diffusion-based text-to-music generation on consumer hardware, it operated primarily as a proof-of-concept. Crucially, v1.0 highlighted the gap between feasible and commercial-grade: acoustic fidelity was bottlenecked by mel-spectrogram representations; semantic consistency fell short of professional standards; and complex editing demands—such as track separation and cover generation—remained unaddressed.

In this work, we present ACE-Step v1.5, a definitive framework designed to bridge the gap between experimental prototypes and production-ready standards. We challenge the conventional trade-off between inference efficiency and generative quality, implementing a refined \textbf{Hybrid Reasoning-Diffusion Architecture} that fundamentally decouples usability from acoustic rendering. Instead of an end-to-end black box, we position the Language Model (LM) as an intelligent Composer Agent and the Diffusion Transformer (DiT~\cite{peebles2023scalablediffusionmodelstransformers, ma2024sitexploringflowdiffusionbased}) as a specialized acoustic renderer. The LM functions as a high-level planner, explicitly reconstructing vague user prompts into precise structural blueprints (e.g., formatting lyrics, calculating duration, and expanding queries) to relieve the DiT of semantic ambiguities. Conversely, the DiT is freed to focus exclusively on acoustic richness and instrument separation. By integrating a novel distillation algorithm, we compress the generation process from 50 steps to merely 4--8 steps, achieving an inference speedup of over $100\times$ while simultaneously enhancing signal-to-noise performance.

Our primary contributions are summarized as follows:

\begin{enumerate}
    \item \textbf{Commercial-Grade Efficiency via Distillation:} We introduce a specialized distillation protocol that compresses the DiT inference trajectory to 4--8 steps. This optimization enables sub-second generation of high-fidelity audio on consumer-grade GPUs ($<$4GB VRAM), effectively democratizing access to professional music synthesis without sacrificing instrumental clarity.
    
    \item \textbf{Decoupled Usability-Generation Architecture:} We propose a structural paradigm shift where the LM and DiT address distinct generative challenges. By offloading structural planning and prompt understanding to the LM via multi-task training, we ensure that the DiT receives standardized, unambiguous conditions, thereby maximizing its generative ceiling.
    
    \item \textbf{Intrinsic RL for Aligned Control:} To enhance controllability without introducing external biases, we employ a Reinforcement Learning pipeline driven by intrinsic rewards derived from the model's own comprehension tasks. This approach ensures robust alignment across over 50 languages and enforces precise adherence to stylistic and lyrical constraints.
    
    \item \textbf{Unified Omni-Task Framework:} Moving beyond simple text-to-music synthesis, ACE-Step v1.5 establishes a unified framework for complex musical manipulation. The model natively supports a versatile toolkit—including cover generation, seamless repainting, track extraction, and vocal-to-BGM conversion—demonstrating that a single, efficiently trained model can satisfy the diverse, long-tail demands of modern music production.
\end{enumerate}
\section{Data Infrastructure}
\label{sec:data_infrastructure}
The unprecedented fidelity and controllability of ACE-Step v1.5 strictly stem from a paradigm shift in data engineering: from quantity-first to alignment-first. We establish a comprehensive Data Infrastructure designed to maximize the information density per training sample.

The core of this infrastructure is a sophisticated Self-Evolving Annotation System. We initially engineered precision prompts to leverage Gemini 2.5 Pro ~\cite{comanici2025gemini25pushingfrontier} —chosen for its world-class reasoning and minimal inherent bias—to annotate a Golden Set of 5 million audio samples. Using this high-quality seed data, we fine-tuned Qwen2.5-Omni to create our proprietary specialist models: ACE-Captioner and ACE-Transcriber. To further eliminate hallucinations and improve adherence, we trained specific Reward Models on 4 million synthesized negative pairs and applied Reinforcement Learning to refine the captioners. Finally, this RL-enhanced pipeline annotated our entire 27-million-sample corpus, followed by a rigorous filtration step.

\begin{figure}[htbp]
    \centering
    \includegraphics[width=0.9\linewidth]{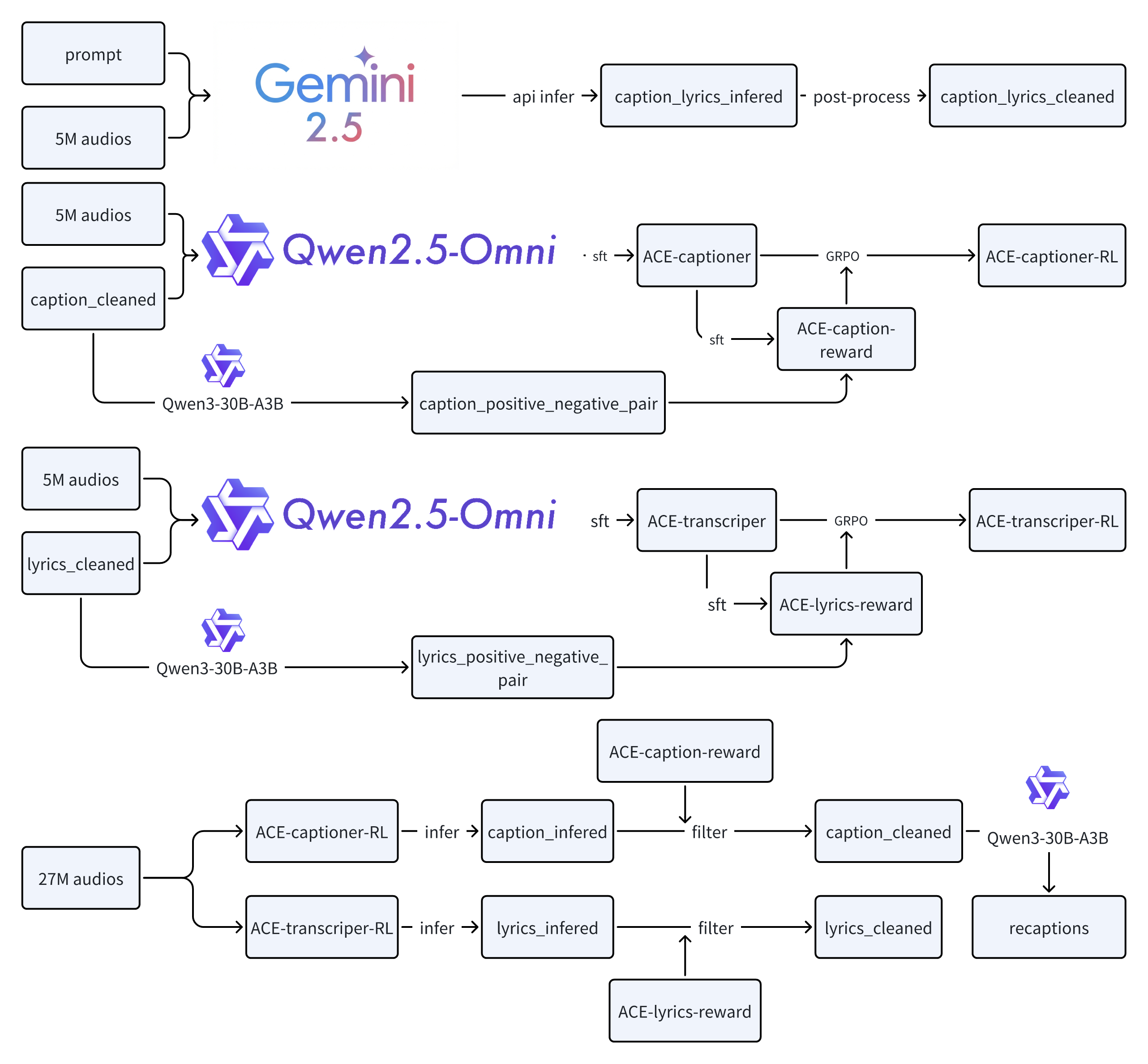}
    \caption{The Data Infrastructure and RL-Driven Annotation Pipeline.}
    \label{fig:data_pipeline}
\end{figure}

\subsection{RL-Driven Annotation Pipeline}
To bridge the gap between acoustic details and textual descriptions, we constructed a multi-stage pipeline that achieves fine granularity in describing over 2,000 musical styles and 50+ languages. 
\textbf{Stage 1 (Foundation SFT)} utilized the 5 million Gemini-annotated samples to supervise-fine-tune Qwen2.5-Omni~\cite{xu2025qwen25omnitechnicalreport}, yielding the base ACE-Captioner and ACE-Transcriber.
\textbf{Stage 2 (Reward Modeling)} involved constructing 4 million contrastive pairs via a dual-track Heuristic Semantic Augmentation strategy. We generated \textit{Hard Negatives} by subtly corrupting semantic elements (e.g., swapping instruments or mood tags) and \textit{Robust Positives} through semantic rewrites. This forces the Reward Models to distinguish fine-grained semantic discrepancies beyond surface-level text patterns.
In \textbf{Stage 3 (RL Refinement)}, we applied GRPO~\cite{shao2024deepseekmathpushinglimitsmathematical} to optimize the captioners against these reward models.
Finally, \textbf{Stage 4} deployed these models to annotate the full 27-million dataset, reusing the Reward Models to filter out low-alignment pairings.

\subsection{Semantic-Acoustic Filtering and Augmentation}
To mitigate the generative instability observed in previous iterations, we implement a rigorous filtering and augmentation protocol driven by the ACE-Reward models. We apply a metric-based filtration strategy where samples exhibiting low audio-text correlation are aggressively discarded, thereby removing noise that contributes to mode collapse. Concurrently, recognizing the distributional shift between our dense training captions and typically sparse user prompts, we employ Qwen3-30B-A3B~\cite{yang2025qwen3technicalreport} for \textit{Query Rewriting}. By augmenting the dataset with diverse input variations—ranging from single-style keywords to summarized descriptions—we ensure the model remains robust across varying prompt complexities and lengths.

\subsection{Multilingual and Timbre Pre-processing}
To facilitate global versatility and precise control, we introduce specialized preprocessing strategies for linguistic and acoustic features. For non-Roman scripts (e.g., Chinese, Japanese, Thai), we implement a stochastic Romanization strategy, converting 50\% of lyrics into phonemic representations during training. This approach enables the model to share phonological representations across languages, significantly enhancing pronunciation accuracy for rare tokens without expanding the vocabulary size. For timbre modeling, we construct a dedicated reference dataset by applying voice masking to isolate vocal or instrumental stems. These segments are concatenated into continuous streams and processed into fixed 30-second context windows with loop padding, enabling the model to learn robust zero-shot timbre cloning capabilities.

\subsection{Progressive Curriculum Learning}
We adopt a three-phase curriculum learning strategy to progressively unlock the model's capabilities using over 27 million processed samples. The training regimen begins with \textbf{Foundation Pre-training} on a massive corpus of approximately 20 million text-to-music pairs, focusing on capturing general acoustic distributions and broad language representations. This is followed by an \textbf{Omni-Task Fine-tuning} phase, where the dataset is refined by retaining the top 50\% of quality samples and augmented with 6 million stem-separated tracks; this phase targets advanced editing tasks such as track extraction and repainting. The pipeline concludes with a \textbf{High-Quality SFT} phase using a curated subset of 2 million samples. Selected via stratified high-reward filtering, this final stage maximizes instruction adherence and stylistic fidelity, ensuring the model aligns strictly with complex user prompts without overfitting to dominant musical forms.
\section{Model Architecture}
\label{sec:arch}

\begin{figure}[htbp]
    \centering
    \includegraphics[width=1.0\linewidth]{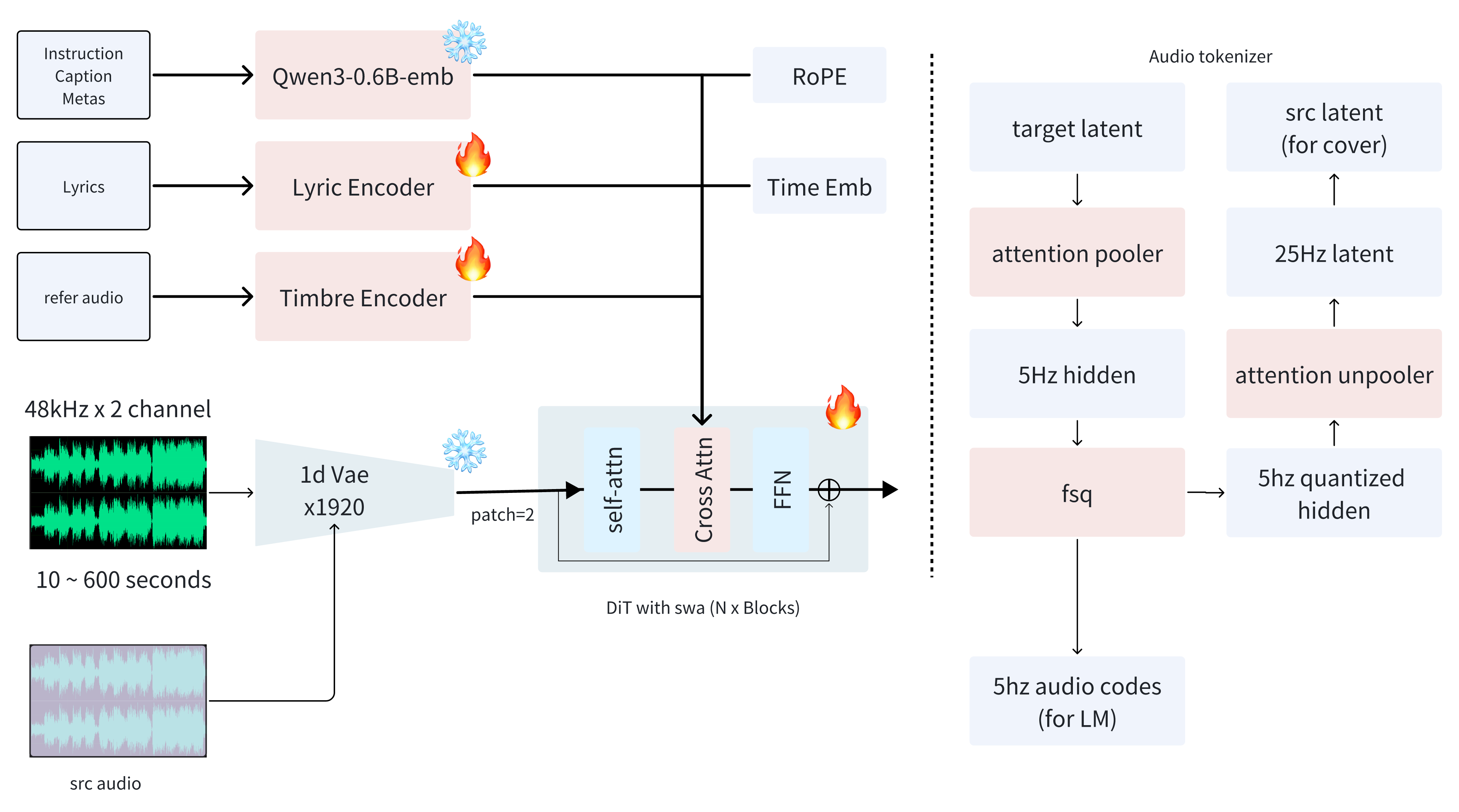}
    \caption{The Model Framework of ACE-Step 1.5.}
    \label{fig:model_framework}
\end{figure}

\subsection{High-Fidelity 1D VAE: Beyond Mel-Spectrograms}
AceStep v1.0 relies on 2D Mel-spectrograms, a representation often plagued by phase loss and insufficient resolution for high-frequency transients such as bass plucks and percussion. To circumvent this acoustic bottleneck, we implement a pure waveform-domain 1D Variational Autoencoder (VAE)~\cite{evans2024stableaudioopen}. Adopting a configuration inspired by SongBloom~\cite{yang2025songbloomcoherentsonggeneration}, our VAE compresses 48kHz stereo audio into a compact 64-dimensional latent space at 25Hz, achieving a substantial $1920\times$ compression rate while preserving near-lossless perceptual quality.

Training is optimized via the Muon optimizer~\cite{liu2025muonscalablellmtraining}, which demonstrates superior convergence properties for massive 1D convolutional layers compared to standard AdamW. The model undergoes adversarial tuning for 600k steps on 120 A100 GPUs. Crucially, in the final 100k steps, we eliminate the KL divergence penalty and increase the adversarial loss weight from 0.1 to 0.5. This strategy significantly sharpens acoustic textures, allowing the 1D VAE to outperform 2D DCAE baselines in reconstruction metrics, particularly regarding vocal clarity and instrumental separation.

\subsection{The DiT Backbone: Efficient Conditioned Generation}
At the core of our acoustic synthesis pipeline lies a Diffusion Transformer (DiT) scaled to approximately 2B parameters. Engineered as a general-purpose conditional generator, this backbone balances computational efficiency with multi-task versatility. To reconcile long-sequence modeling with structural coherence, we incorporate a Hybrid Attention mechanism: odd layers utilize Sliding Window Attention to capture local acoustic nuances and transients, while even layers employ Global Group Query Attention (GQA) to maintain long-term rhythmic and melodic consistency.

The model operates on a unified conditioning stack where Qwen3-0.6B caption embeddings are concatenated with dedicated timbre and lyric encoders and injected via Cross-Attention. To bridge the continuous latent space with our discrete LM planner, we implement a Finite Scalar Quantization (FSQ~\cite{mentzer2023finitescalarquantizationvqvae}) tokenizer. This module utilizes attention pooling to compress 25Hz latents into 5Hz discrete codes (Codebook $\approx$ 64k), serving as the structural Source Latent. Finally, the model processes a composite input tensor—combining Source, Noised Target, and Mask—through a patchify layer, effectively halving the sequence length to 12.5Hz for optimal throughput.

\begin{figure}[htbp]
    \centering
    \includegraphics[width=1.0\linewidth]{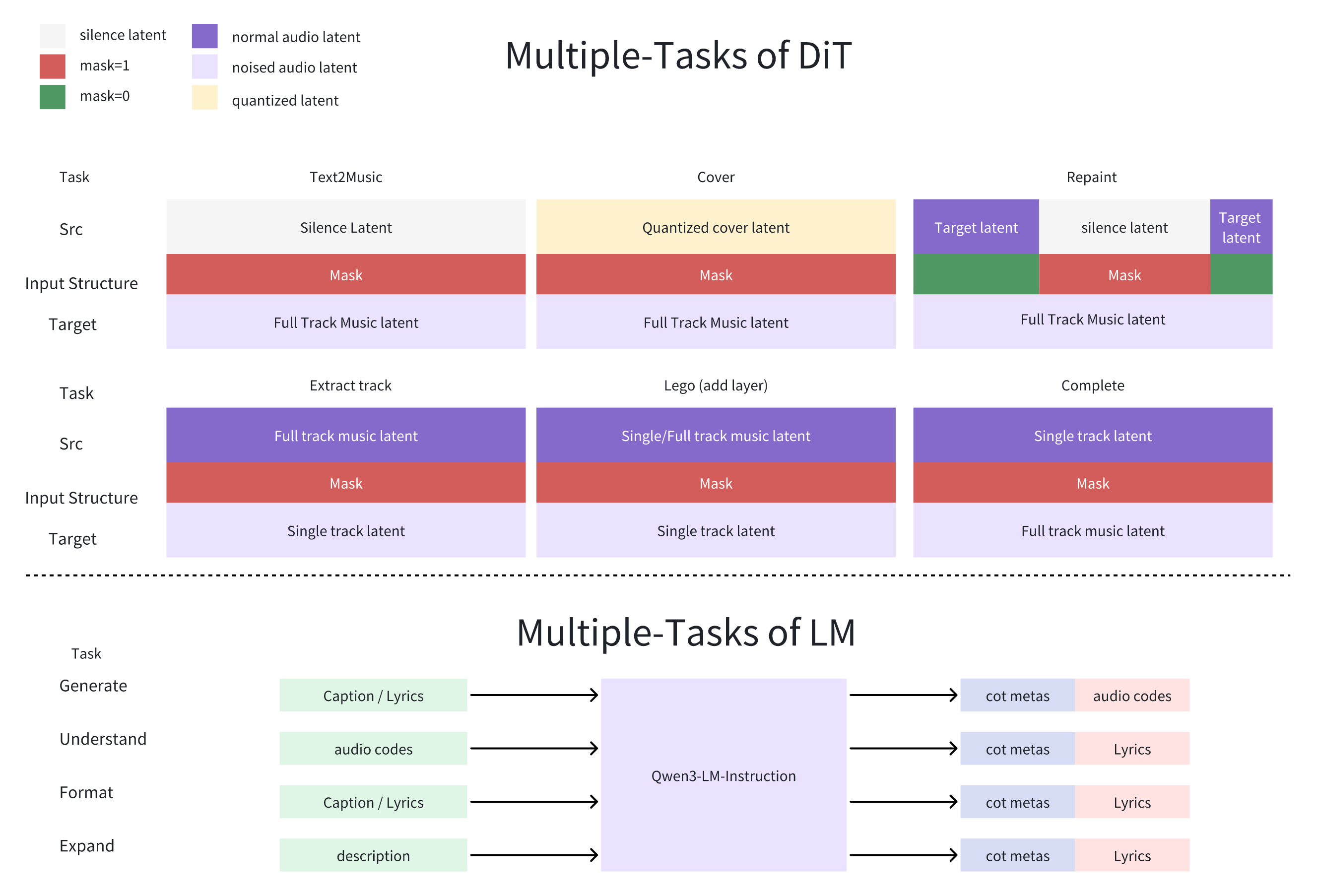}
    \caption{The Multi-Modal Applications of ACE-Step 1.5 enabled by the Masked Generative Framework.}
    \label{fig:application_map}
\end{figure}

\subsection{Omni-Task Formulation}
We unify music generation into a flexible Masked Generative Framework, where manipulating the \textit{Source Latent} and \textit{Mask} configuration allows a single model to support six distinct modalities. Beyond standard \textbf{Text-to-Music}, the architecture enables \textbf{Cover Generation} by re-synthesizing timbre while retaining melodic skeletons via quantized latents, and \textbf{Repainting} for seamless segment regeneration. Furthermore, it generalizes to complex structural tasks, including \textbf{Track Extraction}, \textbf{Layering} (adding complementary instruments to a track), and \textbf{Completion} (orchestrating full arrangements around a single motif). This approach not only streamlines the training process by leveraging a shared latent space but also enhances efficiency, as the model can adapt to diverse inputs without requiring task-specific fine-tuning. By incorporating quantized latents, the framework ensures high-fidelity preservation of musical elements like melody and rhythm during transformations, making it ideal for creative remixing applications.

\subsection{The Language Model: Reasoning and Planning}
While the DiT handles acoustic rendering, logical planning is delegated to a Qwen-based Language Model. Utilizing the ChatML template, the LM is trained to function as a ``Composer Agent,'' generating structured Chain-of-Thought (CoT) metadata—including BPM, key, duration, and structure—in YAML format prior to outputting content. This architecture enables four distinct interaction paradigms: (1) \textbf{Planner Mode}, where the model translates vague user prompts into specific structural blueprints and audio codes; (2) \textbf{Listener Mode}, demonstrating semantic comprehension by reverse-engineering captions and lyrics from audio codes; (3) \textbf{Co-Pilot Mode}, acting as a creative assistant to expand simple queries into full song structures; and (4) \textbf{Refiner Mode}, where the LM standardizes and optimizes stylized or raw user inputs to maximize the downstream DiT's performance. The integration of CoT reasoning allows the LM to break down complex musical ideas into logical steps, improving the coherence and artistic quality of generated outputs. Furthermore, by outputting metadata in a standardized YAML format, it facilitates seamless interoperability with other tools or systems in music production pipelines.
\section{Training Strategy}
\label{sec:training}

\begin{figure}[htbp]
    \centering
    \includegraphics[width=1.0\linewidth]{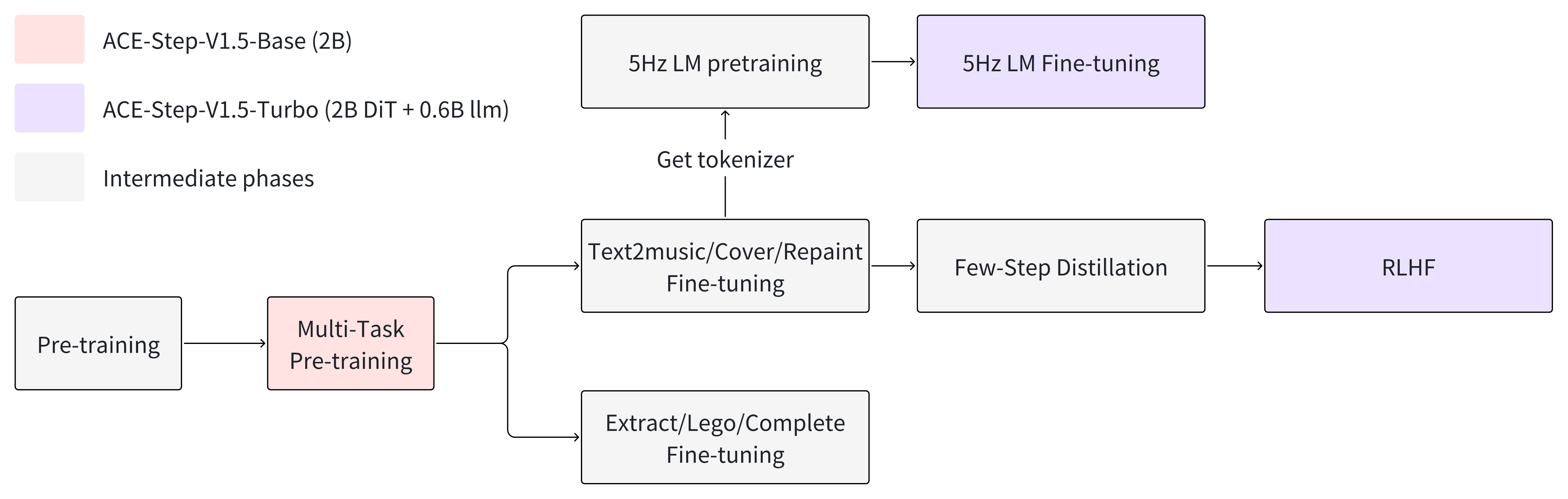}
    \caption{The training pipeline of ACE-Step v1.5.}
    \label{fig:model_zoo}
\end{figure}

We adopt a progressive training paradigm evolving from foundational generative priors to complex multi-task fine-tuning, culminating in adversarial distillation that enables the student model to surpass its teacher. The initial generative pre-training phase uses a 20M-sample foundation dataset (Source = Silence) to establish robust priors for musical structure and acoustic distribution. The subsequent omni-task mixture phase introduces the Omni-Task Protocol with 17M samples, using varying Source and Mask conditions to decouple stylistic attributes from melodic content for precise editing tasks. The final high-fidelity supervised fine-tuning (SFT) phase employs a curated 2M-sample high-alignment subset, shifting the objective from diversity to maximizing correlation between control signals and outputs for robust initialization.

\subsection{Adversarial Dynamic-Shift Distillation}

To enable sub-second inference without quality degradation, we employ a distillation framework based on Decoupled DMD2~\cite{yin2024improveddistributionmatchingdistillation, liu2025decoupleddmdcfgaugmentation}, augmented with a GAN objective and a ConvNeXt-based discriminator~\cite{novack2025fasttexttoaudiogenerationadversarial} operating in the latent space via Flow Matching objectives. To overcome limitations where fixed time-step shifts restrict denoising trajectory diversity, we propose a dynamic-shift strategy with the shift parameter stochastically sampled from $\{1, 2, 3\}$. This exposes the surrogate model to diverse denoised states, preventing overfitting and ensuring stable adversarial feedback. The distilled model reduces inference from 50 to 8 steps (Shift=3) without Classifier-Free Guidance, achieving $200\times$ speedup to generate 240-second tracks in $\sim$1 second on an NVIDIA A100. The student outperforms the teacher by mitigating discretization errors and sharpening textures via the discriminator, facilitating test-time scaling for autonomous generation and re-ranking of candidate batches using intrinsic rewards.

\subsection{Self-Supervised Intrinsic Alignment}
\label{sec:intrinsic_rl}

To enhance controllability without external bias, we establish a unified intrinsic reinforcement learning framework, applying DiffusionNTF~\cite{zheng2025diffusionnftonlinediffusionreinforcement} to the DiT backbone and Group Relative Policy Optimization (GRPO) to the Language Model, deriving rewards from internal geometric and probabilistic consistency. For DiT alignment, we hypothesize that low-fidelity generation correlates with high entropy in cross-attention maps and develop the Attention Alignment Score (AAS), measuring consensus between Token-to-Frame and Frame-to-Token attention maps. AAS aggregates three properties via Dynamic Time Warping (DTW): (1) coverage, the ratio of lyric tokens with significant energy peaks; (2) monotonicity, penalizing non-forward alignment of attention centroids; and (3) path confidence, the average energy density along the alignment trajectory. Optimizing AAS reduces hallucinations and achieves $>95\%$ correlation with human judgments for lyric-audio synchronization. For LM alignment, we treat the model as both a ``Composer'' (Text-to-Code) and a ``Listener'' (Code-to-Text), constructing a reward model using Pointwise Mutual Information (PMI) for semantic adherence. Unlike raw likelihood, PMI penalizes generic descriptions and rewards captions specific to the generated audio codes. The final reward is a dynamic weighted sum emphasizing stylistic atmosphere (50\%), lyrical content (30\%), and metadata constraints (20\%), guiding the LM to prioritize artistic nuance while maintaining structural validity.
\section{Evaluation}
\label{sec:evaluation}

We conduct a comprehensive evaluation across three tiers: objective metrics, subjective human preference, and a novel usability assessment targeting real-world deployment scenarios.

\subsection{Objective Metrics}
\label{sec:obj_metrics}

We evaluate on a diverse test set spanning 20 musical styles and bilingual (Chinese/English) prompts, using Gemini 3 Pro to generate structured prompts and corresponding lyrics.

\textbf{Metric Sources.} AudioBox scores derive from Meta's aesthetic assessment framework ~\cite{tjandra2025metaaudioboxaestheticsunified}; SongEval metrics are computed via the ASLP-lab evaluation suite ~\cite{yao2025songevalbenchmarkdatasetsong}. Style and Lyric Alignment are quantified using our proprietary ACE-Caption-Reward and ACE-Lyric-Reward models.

\bgroup
\def\arraystretch{1.1} 
\begin{table}[t]
\centering
\caption{Comparison with commercial (top) and open-source (bottom) music generation models. The evaluation configuration is 1.7B LM + dynamic distilled DiT (no RL). \textbf{Bold} = best, \underline{underline} = second best. $\uparrow$ higher is better.}
\label{tab:comparison}
\resizebox{\textwidth}{!}{
\begin{tabular}{lccccccccccc}
\toprule
\multirow{2}{*}{\textbf{Model}} & \multicolumn{4}{c}{\textbf{AudioBox} $\uparrow$} & \multicolumn{5}{c}{\textbf{SongEval} $\uparrow$} & \multirow{2}{*}{\textbf{\begin{tabular}[c]{@{}c@{}}Style\\Align $\uparrow$\end{tabular}}} & \multirow{2}{*}{\textbf{\begin{tabular}[c]{@{}c@{}}Lyric\\Align $\uparrow$\end{tabular}}} \\
\cmidrule(lr){2-5} \cmidrule(lr){6-10}
 & CE & CU & PC & PQ & Coh. & Mus. & Mem. & Cla. & Nat. &  &  \\
\midrule
Udio-v1.5~ & 7.45 & 7.65 & 6.15 & 8.03 & 4.15 & 3.96 & 4.09 & 3.93 & 3.86 & 34.9 & 24.8 \\
Suno-v4.5~ & 7.63 & 7.85 & 6.22 & 8.25 & 4.64 & 4.51 & 4.63 & 4.53 & 4.49 & 40.5 & \underline{32.7} \\
Suno-v5~ & \underline{7.69} & 7.87 & \textbf{6.51} & 8.29 & \textbf{4.72} & \underline{4.62} & \underline{4.71} & \underline{4.63} & \underline{4.56} & \textbf{46.8} & \textbf{34.2} \\
Mureka-V7.6~ & 7.44 & 7.71 & 6.35 & 8.13 & 4.43 & 4.29 & 4.35 & 4.29 & 4.21 & 36.2 & 22.4 \\
MinMax-2.0~ & \textbf{7.71} & \underline{7.95} & 6.42 & \textbf{8.38} & 4.61 & 4.51 & 4.59 & 4.50 & 4.41 & \underline{43.1} & 29.5 \\
\midrule
Yue ~\cite{yuan2025yuescalingopenfoundation} ~ & 6.58 & 7.29 & 4.95 & 7.39 & 3.01 & 2.80 & 2.85 & 2.79 & 2.82 & 26.8 & $-$4.6 \\
ACE-Step 1.0 ~ & 7.22 & 7.52 & \underline{6.50} & 7.76 & 3.99 & 3.73 & 3.85 & 3.78 & 3.68 & 28.5 & 0.9 \\
LeVo ~\cite{lei2025levohighqualitysonggeneration}~ & 7.61 & 7.78 & 5.92 & 8.31 & 3.55 & 3.35 & 3.32 & 3.31 & 3.20 & 29.4 & $-$1.2 \\
DiffRhythm 2 ~\cite{jiang2025diffrhythm2efficienthigh}~ & 7.25 & 7.61 & 6.33 & 7.99 & 3.99 & 3.79 & 3.97 & 3.82 & 3.66 & 32.1 & 3.8 \\
HeartMuLa~ ~\cite{yang2026heartmulafamilyopensourced}& 7.66 & 7.89 & 6.15 & 8.25 & \underline{4.68} & 4.55 & 4.69 & 4.55 & 4.45 & 31.7 & 28.6 \\
\midrule
ACE-Step 1.5 & 7.42 & \textbf{8.09} & 6.47 & \underline{8.35} & \textbf{4.72} & \textbf{4.67} & \textbf{4.72} & \textbf{4.66} & \textbf{4.59} & 39.1 & 26.3 \\
\bottomrule
\end{tabular}%
}
\end{table}
\egroup

\textbf{Results.} As shown in Table~\ref{tab:comparison}, ACE-Step 1.5 achieves the highest AudioBox CU (\textbf{8.09}), second-best PQ (\underline{8.35}), and tied-best Coherence (\textbf{4.72}), while demonstrating superior temporal consistency (Mem \textbf{4.72}, Nat \textbf{4.59}). While commercial models like Suno-v5 maintain advantages in style/lyric alignment (46.8/34.2), ACE-Step 1.5 shows strong competitiveness against open-source baselines (39.1/26.3), validating that our distillation preserves high-fidelity acoustic rendering.

\subsection{Subjective Evaluation via Music Arena}
\label{subsec:arena}

To assess perceptual quality beyond automated metrics, we adopt the Music Arena protocol. Human evaluators perform blind A/B tests on pairwise model outputs given identical prompts/lyrics. Aggregating Bradley-Terry scores, ACE-Step 1.5's subjective quality ranks between Suno-v4.5 and Suno-v5, confirming that sub-second generation does not compromise aesthetic fidelity.

\subsection{Usability: A New Evaluation Paradigm}
\label{sec:usability}

We argue that prevailing benchmarks inadequately capture the practical utility of text-to-music systems in production workflows. Consequently, we propose \textbf{Usability} as the holistic efficacy metric, formalized via a rigorous 17-point Checklist across six critical dimensions (Table~\ref{tab:usability}).

\begin{table*}[t] 
\centering
\small
\caption{The 17-point Usability Checklist for evaluating text-to-music systems in real-world creative workflows.}
\label{tab:usability}
\begin{tabular}{@{}clp{4.2cm}p{7.5cm}@{}}
\toprule
\textbf{Dimension} & \textbf{ID} & \textbf{Criterion} & \textbf{Requirement / Success Condition} \\
\midrule
\multirow{3}{*}{\parbox{2.8cm}{\raggedright 1. Democratization \& Access}} 
& 1 & Deployment Barrier & Must run on consumer-grade local hardware, not gated behind enterprise APIs/clusters. \\
& 2 & Hardware Agnosticism & Operability on widespread GPUs (e.g., $<$8GB VRAM), without mandating specialized H100s. \\
& 3 & Installation Complexity & User-ready packaged solution (e.g., pip/conda installable), not a fragmented research repository. \\
\midrule
\multirow{3}{*}{\parbox{2.8cm}{\raggedright 2. Efficiency \& The ``Surprise Factor''}} 
& 4 & Time-to-Result & Generation latency must not break the user's creative ``flow state'' (sub-second to few seconds). \\
& 5 & Test-Time Scaling & Inference speed supports massive parallel batching for efficient latent space exploration. \\
& 6 & The Serendipity Coefficient & Capability to rapidly produce high-volume diverse candidates to maximize ``happy accidents''. \\
\midrule
\multirow{3}{*}{\parbox{2.8cm}{\raggedright 3. Robustness \& Scale}} 
& 7 & Prompt Robustness & Resilience to sparse/vague inputs (the ``Anti-Gacha'' requirement: no mode collapse on weak prompts). \\
& 8 & Temporal Scalability & Seamless handling from 10-second sound design loops to 10-minute progressive compositions. \\
& 9 & Linguistic Breadth & Coverage of the long tail of global languages ($>$50), avoiding Anglo-centric bias. \\
\midrule
\multirow{3}{*}{\parbox{2.8cm}{\raggedright 4. Professional Integration}} 
& 10 & Non-Destructive Editing & Surgical ``Repaint'' capabilities allowing partial edits without regenerating the entire track. \\
& 11 & Stem Separation & Output of isolated tracks (Vocals/Drums/Bass/Other) for downstream mixing and mastering. \\
& 12 & Reference Control & Zero-shot timbre cloning and structural guidance from arbitrary audio references. \\
\midrule
\multirow{3}{*}{\parbox{2.8cm}{\raggedright 5. Creative Symbiosis}} 
& 13 & Inspiration Unblocking & Function as a ``motif generator'' to jumpstart creativity during writer's block. \\
& 14 & Identity Consistency & Ability to maintain a consistent sonic identity across multiple generations/consecutive clips. \\
& 15 & Exploratory Playability & Interface allowing fluid toggling between structured planning and unstructured jamming. \\
\midrule
\multirow{2}{*}{\parbox{2.8cm}{\raggedright 6. Knowledge \& Precision}} 
& 16 & Professional Terminology & Strict adherence to technical instructions (e.g., ``Sidechain Compression,'' ``Phrygian Mode,'' ``TB-303 Acid Line''), translating jargon into accurate acoustic phenomena rather than generic vibes. \\
& 17 & World Knowledge Grounding & Broad cultural literacy for specific eras, regional styles, and artist references (e.g., ``1920s Shanghai Jazz,'' ``Cyberpunk 2077 Soundscape'') without anachronistic hallucination. \\
\bottomrule
\end{tabular}
\end{table*}

This framework shifts evaluation focus from isolated quality scores to \textit{system capability} in professional creative ecosystems, addressing deployment barriers and workflow integration that standard benchmarks often overlook.
\newpage
\section{Conclusion}
\label{sec:conclusion}

In this work, we presented ACE-Step v1.5, a definitive framework that democratizes commercial-grade music generation through a novel Hybrid Reasoning-Diffusion Architecture. By decoupling structural planning from acoustic rendering, we achieved sub-second inference on consumer hardware without compromising fidelity, enabling a unified toolkit for synthesis, cover generation, and precise editing. This approach effectively bridges the gap between open-source capabilities and proprietary systems, offering a versatile foundation for creative workflows.

However, current capabilities remain bounded by the parameter constraints necessary for consumer-grade accessibility. Future development will focus on scaling model size and deepening our data infrastructure to address these limitations. Specifically, we aim to enhance acoustic richness, expand support for diverse languages and musical styles, and integrate broader world knowledge for professional-grade caption control. Furthermore, we plan to refine precise lyric alignment and enable agentic audio editing workflows. We believe that continued scaling, combined with our rigorous data engineering pipeline, will resolve these constraints and further push the boundaries of generative music.

\newpage
\section*{Acknowledgements}
\label{sec:acknowledgements}

This project is co-led by ACE Studio and StepFun.

We extend our profound gratitude to StepFun for their generous provision of computing power and storage infrastructure. Without such critical support, the development of ACE-Step would not have been possible. We are also deeply thankful to their team for organizing and conducting the subjective evaluations of our model.

We are immensely grateful to Jing Guo and Wenxiao Zhao for their intensive involvement in decision-making and discussions concerning the training and application aspects of ACE-Step. Their meticulous weighing of trade-offs was instrumental in shaping ACE-Step into the versatile foundation model it has become.

Our sincere thanks go to Sen Wang and Shengyuan Xu for their invaluable support in data engineering. Effective data management was foundational to our efforts, and ACE-Step could not have been successfully trained without their expertise. We also thank Yulin Song for his collaboration on the data pipeline construction and his specific contributions to the reinforcement learning reward framework.

We wish to express our deepest appreciation to Junmin Gong for his extensive contributions, which encompassed algorithm development, framework design, conducting experiments, model training, and evaluation. Furthermore, we thank him for authoring the report, preparing the open-source code release, and creating the project webpage.

Finally, we extend a special thank you to the ACE-Step community for their invaluable feedback and resource sharing. ACE-Step is a community-driven model, and we could not have come this far without the continuous engagement and support of our users.

\newpage
\bibliographystyle{unsrt}
\bibliography{references}  
\end{document}